\documentstyle[12pt]{article}
\begin{document}
\begin{flushright}
Alberta Thy-31-94\\October, 1994

\end{flushright}

\begin{center}
{\Large \bf Nonfactorization and Color-Suppressed $
B\rightarrow\psi(\psi(2S))+K(K^*) $
Decays}\\[10mm]
A. N. Kamal and A. B. Santra\footnote{on leave of absence from Nuclear Physics
Division,
Bhabha Atomic Research Centre, Bombay-400085,  India.}\\[5mm]
{\em Theoretical Physics Institute and Department of Physics,\\ University of
Alberta,
Edmonton, Alberta T6G 2J1, Canada.}
\end{center}
\vskip1cm

\begin{abstract}
Using $N_c=3$ value of the parameter $a_2=0.09$ but including a modest
nonfactorized
amplitude, we show that it is possible to understand all data, including
polarization, for
color-suppressed $B\rightarrow\psi(\psi(2S))+K(K^*) $ decays in all commonly
used models
of form factors. We show that for $B\rightarrow\psi +K $ decay one can define
an effective $
a_2$, which is process-dependent and, in general, complex; but it is not
possible to define
an effective  $a_2$ for  $B\rightarrow\psi +K^* $ decay. We also explain why
nonfactorized
amplitudes do not play a significant role in color-favored B decays.
\end{abstract}

\newpage

It was shown, in ref. \cite{gkp}, that within the factorization approximation,
the commonly
used models for $B \rightarrow K(K^*) $ transition form factors failed to
account
simultaneously for the following two measured ratios,
\begin{eqnarray}
R \equiv { \Gamma(B \rightarrow \psi K^*) \over \Gamma(B \rightarrow \psi K)},
\nonumber\\
P_L \equiv { \Gamma_L(B \rightarrow \psi K^*) \over \Gamma(B \rightarrow \psi
K^*)}.
\end{eqnarray}
In this note, we have relaxed the factorization approximation to allow
nonfactorized
contributions to the decay amplitudes and demonstrated that all the commonly
used models
for the transition form factors can be consistent not only with the quantities
R and $ P_L$ of
eqn.  (1)  but also with the following three quantities \cite{ks}
\begin{eqnarray}
R' \equiv { \Gamma(B \rightarrow \psi K) \over \Gamma(B \rightarrow \psi' K)},
\nonumber\\
R^{\prime\prime}\equiv { \Gamma(B \rightarrow \psi K^*) \over \Gamma(B
\rightarrow \psi'
K^*)},
\end{eqnarray}
and the measured value of $B(B \rightarrow \psi K) $ \cite{cleo,pdg}. Here $
\psi'$ is $
\psi(2S)$.

          We begin with some definitions relevant to the analysis of $ B
\rightarrow \psi (\psi') +
K(K^*) $. The relevant part of the weak Hamiltonian for $ b \rightarrow
c\bar{c} s$ decay is

  \cite{nrsx},
\begin{eqnarray}
H_w = {G_F \over \sqrt{2} } V_{cb} V_{cs}^* \left\{ C_1(\bar{c}b) (\bar{s}c) +
C_2 (\bar{c}c)
(\bar{s}b) \right\}.
\end{eqnarray}
Here $ (\bar{c}b)$ etc. represent color-singlet (V-A) brackets and $C_1$ and
$C_2$ are the
Wilson coefficients for which several values can be found in the literature:
$C_1=1.12$,
$C_2=-0.26$  \cite{nrsx,bshf1};  $C_1=1.13$,   $C_2=-0.29$  \cite{sfm}. We
adopt the
values,  $C_1=1.12\pm0.01$,  $C_2=-0.27\pm0.03$. Fierz-transforming the
color-singlet
combinations in eqn.  (3) in color space, we obtain, with number of colors $
N_c = 3$,
\begin{eqnarray}
(\bar{c}b)(\bar{s}c) = {1 \over 3} (\bar{c}c)(\bar{s}b) + {1 \over 2}
\sum_{a=1}^{8}{(\bar{c}\lambda^ab)(\bar{s}\lambda^ac)} ,
\end{eqnarray}
where $ \lambda^a$ are the Gell-Mann matrices. $ {1 \over 2}
\sum{(\bar{c}\lambda^ab)(\bar{s}\lambda^ac)}$  $ (\equiv H_w^{(8)})$ being a
product of
two color-octet currents contributes to the nonfactorized part of the decay
amplitude.

          The  amplitudes for $B \rightarrow \psi K(K^*)$  decays can be
written using eqns.  (3)
and (4) as,
\begin{eqnarray}
A(B \rightarrow \psi K(K^*)) ={G_F \over \sqrt{2} } V_{cb} V_{cs}^* a_2\left[
\left\langle \psi
K(K^*) \mid (\bar{c}c)(\bar{s}b)\mid B \right\rangle \right. \nonumber\\
\left.+ \kappa \left\langle \psi K (K^*)\mid H_w^{(8)}\mid B
\right\rangle\right],
\end{eqnarray}
 where $a_2 = C_2 + C_1 /3 = 0.10\pm0.03$ and $\kappa=C_1 / a_2.$ The first
term in eqn.
(5) can be evaluated using factorization procedure  \cite{bsw}. The second term
accounts
for the nonfactorization contribution.  Since $ \kappa$ is large, of the order
of ten, even a
small amount of nonfactorized  contribution will have a significant effect on
the amplitudes.
We recognize that there could be nonfactorized contributions to the first term
on the right
hand side of eqn.  (5), however, we anticipate the (nonfactorized) contribution
of the second
term to dominate due to the largeness of  $\kappa $.    We write the Lorentz
structures of
$\left\langle \psi K (K^*)\mid H_w^{(8)}\mid B \right\rangle $, for ease of
comparison with the
factorized amplitude, as
\begin{eqnarray}
{\left\langle K \psi \mid H_w^{(8)}\mid B \right\rangle}&=&m_\psi
f_\psi(\varepsilon_1^*.p_B)
F_1^{NF}(m_\psi^2), \nonumber \\
{\left\langle K^* \psi \mid H_w^{(8)}\mid B \right\rangle}&=&m_\psi
f_\psi\left[ (m_B +
m_{K^*})(\varepsilon_1^*.\varepsilon_2^*)A_1^{NF} \right. \nonumber \\
&-&\left. {(\varepsilon_2^*.(p_B  - p_{K^*}))(\varepsilon_1^*.(p_B +
p_{K^*}))\over (m_B +
m_{K^*})}A_2^{NF} \right. \nonumber \\
&+&\left. {2i\over (m_B + m_{K^*})} \varepsilon_1^{*\mu} \varepsilon_2^{*\nu}
\varepsilon_{\mu\nu\alpha\beta}p_{K^*}^\alpha p_B^\beta V^{NF}   \right],
\end{eqnarray}
where $ \varepsilon_1$ and $ \varepsilon_2$ are the polarization vectors of $
\psi$ and $
K^*$ respectively. The factorized part of the amplitude is obtained by
replacing $
F_1^{NF}$, $ A_1^{NF}$, $ A_2^{NF}$ and $ V^{NF}$ by $ F_1^{BK}$,  $
A_1^{BK^*}$, $
A_2^{BK^*}$ and $ V^{BK^*}$, the relevant form factors  \cite{bsw}
respectively.

         We make here one very plausible assumption: In $ B \rightarrow \psi
K^*$ decay the
nonfactorizable amplitude contributes only to S-wave final states. This implies
that we retain
$ A_1^{NF}$ which contributes to S-wave but neglect  $A_2^{NF}$ (D-wave) and  $
V^{NF}$(P-wave). The rationale for this assumption is that the t- and u-channel
exchanges
in $ H_w^{(8)} + B \rightarrow \psi + K^*$ involve particles at least as heavy
as the b-flavor
(${\ \lower-1.2pt\vbox{\hbox{\rlap{$>$}\lower5pt\vbox{\hbox{$\sim$}}}}\ }$5 GeV
and as the
momentum in the reaction is $\approx$ 1.5 GeV, it is hard to produce higher
partial waves
through four-point functions. S-waves, if allowed, would dominate.  We
emphasize that the
factorized amplitude is immune to these arguments.

          With our definition of  factorized and nonfactorized amplitudes, we
evaluate $A(B
\rightarrow \psi (\psi') + K(K^*))$, and  write the expressions for $ R$, $
P_L$, $ R^\prime$, $
R^{\prime\prime}$ and $B(B\rightarrow\psi K)$ as follows  \cite{gkp,ks}:

\begin{eqnarray}
R=1.082 \left[ {A_1^{BK^*}(m_\psi^2) \over F_1^{BK}(m_\psi^2)} \right] ^2
{(a\xi - b x)^2 +
2(\xi^2 + c^2 y^2) \over \eta^2},
\end{eqnarray}
\begin{eqnarray}
P_L&=&{(a\xi - b x)^2 \over (a\xi - b x)^2 + 2(\xi^2 + c^2 y^2)},
\end{eqnarray}
\begin{eqnarray}
R'&=&(4.178\pm0.515) \left[ {F_1^{BK} (m_\psi^2)\over F_1^{BK} (m_{\psi'}^2)}
\right]^2
{\eta ^2\over \eta^{'2}},
\end{eqnarray}
\begin{eqnarray}
R^{\prime\prime}&=&(1.845\pm0.227)\left[ {A_1^{BK^*} (m_\psi^2)\over A_1^{BK^*}
(m_{\psi'}^2)} \right]^2{(a\xi - b x)^2 + 2(\xi^2 + c^2 y^2) \over (a'\xi' - b'
x')^2 + 2(\xi^{'2} +
c^{'2} y^{'2})},
\end{eqnarray}
\begin{eqnarray}
B(B \rightarrow \psi K)&=&(2.63\pm 0.19)  \mid a_2 \eta
F_1^{BK}(m_\psi^2)\mid^2 \%,
\end{eqnarray}
where,
\begin{eqnarray}
\xi&=&1 + \kappa{A_1^{NF}(m_\psi^2) \over A_1^{BK^*}(m_\psi^2)},\nonumber \\
\eta&=&1 + \kappa{F_1^{NF}(m_\psi^2) \over F_1^{BK}(m_\psi^2)},\nonumber \\
x&=&{A_2^{BK^*} (m_\psi^2)\over A_1^{BK^*} (m_\psi^2)}, \\
y&=&{V^{BK^*} (m_\psi^2)\over A_1^{BK^*} (m_\psi^2)} ,\nonumber\\
a&=&{m_B^2 - m_{K^*}^2 - m_\psi^2 \over 2m_{K^*}m_\psi},\nonumber\\
b&=&{2\mid\vec{p}_{\psi K^*}\mid^2m_B^2 \over m_{K^*}m_\psi\left( m_B + m_{K^*}
\right)^2} ,\nonumber\\
c&=&{2\mid\vec{p}_{\psi K^*}\mid m_B \over \left( m_B + m_{K^*}
\right)^2}.\nonumber
 \end{eqnarray}
The primed quantities in eqns.  (9) and (10) are obtained from the unprimed
ones by
replacing $ \psi$  with $ {\psi'}$. We have used $ V_{cb} = 0.04$ and $ \tau_B
=
1.5\times10^{-12}s$
  \cite{pdg} in determining eqn.  (11). We have also used  \cite{nrsx}:
$f_\psi=(384\pm14)$
MeV and $f_{\psi'}=(282\pm14)$ MeV. The errors in eqns.  (9), (10) and (11)
reflect the
errors in $f_\psi$ and $f_{\psi'}$.

It is evident from the above (see eqn. (11)) that for $ B\rightarrow \psi K$
decay, involving a
single Lorentz scalar,  it is possible to define an effective $a_2$, $a_2^{eff}
= a_2\eta$;
however,  it is not possible to define an effective $a_2$ for $B\rightarrow
\psi K^*$ as this
amplitude involves three independent Lorentz scalars.

Now, to the experimental data. For the ratio R we use  \cite{gkp,cleo},
\begin{eqnarray}
R_{expt} = 1.71\pm0.34.
\end{eqnarray}
For  $ P_L$ we take the weighted average of three measurements:
$0.80\pm0.08\pm0.05$
\cite{cleo}, $ 0.66\pm0.10_{-0.08}^{+0.10}$  \cite{cdf} and $
0.97\pm0.16\pm0.15$
\cite{argus},
\begin{eqnarray}
P_L = 0.78\pm0.07.
\end{eqnarray}
{}From ref.  \cite{pdg} we calculate,
\begin{eqnarray}
R^\prime_{expt} = 1.48 \pm 0.46, \nonumber \\
R^{\prime\prime}_{expt} = 1.13 \pm 0.50.
\end{eqnarray}
We emphasize that the error assignments are ours, where we have reduced the
propagated
error by one-third assuming that some of the systematic errors would cancel in
the ratio.
For $B(B \rightarrow \psi K)$ we use the weighted average of $B(B^+ \rightarrow
\psi K^+)$
and $B(B^0 \rightarrow \psi K^0)$  \cite{pdg},
\begin{eqnarray}
B(B \rightarrow \psi K) = (0.094 \pm 0.012 )\%.
\end{eqnarray}

In our description there are four parameters, $ \xi$, $ \eta$, $ \xi'$ and $
\eta'$. Eventually,
we reduce them to three by a particular choice of eqn. (17) in the following. $
x$ and  $y$
are not free parameters; rather their allowed range is determined by the
experimental value
of $ P_L$ as detailed below.

In Fig. (1) we have plotted the range of the ratios $x$ and $y$ (see eqn. (12))
allowed by the
polarization data of eqn. (14) for different values of $\chi$  $(=
A_1^{NF}(m_\psi^2) / $
$A_1^{BK^*}(m_\psi^2))$ and the values of  $C_1$ and  $C_2$ (equivalently $a_1$
and $
a_2$) shown in the figure caption. We note that the predictions of all the
models considered
in ref.  \cite{gkp} become consistent with the polarization data with $\chi
\approx 0.12 $, a
value which is eminently plausible.

Next, we calculate $ P_L$, R, $ R^\prime$ and  $R^{\prime\prime}$ and $B(B
\rightarrow
\psi K) $ in six representative models (see ref.  \cite{gkp} for details): BSWI
 \cite{bsw},
where all form factors are calculated at $ q^2=0$ and extrapolated with
monopole forms;
BSWII  \cite{gkp,nrsx}, where $ A_1^{BK^*}$ has a monopole extrapolation but $
F_1^{BK}$,
$A_2^{BK} $ and  $V^{BK^*}$ have dipole behavior; CDDFGN  \cite{cddfgn}, where
the
heavy to light transition form factors are calculated at zero recoil and
extrapolated with
monopole forms; HSQ  \cite{hsq}, where the strange quark is treated as heavy
and the form
factors are extrapolated from $ q^2 = q_{max}^2$ to $ m_\psi^2$ by the method
described in
ref.  \cite{nrsx}; JW  \cite{jw}, where the form factors are calculated at $
q^2 = 0$ in a
light-front formalism and extrapolated to $ m_\psi^2$ using a particular
two-parameter
formula; and IW scheme  \cite{gkp,iw} where form factors measured in $ D
\rightarrow
K(K^*)$ semileptonic decays are continued to $ B \rightarrow K(K^*)$
transitions.  We wish
to emphasize that the  ``experimental'' determination of the form factors in $
D \rightarrow
K(K^*)$ transitions are not model-free as a monopole assumption is made for all
form
factors.

In Table (1)  we have shown a sampling of successful predictions  for all the
measured
quantities in these models. In this Table we have introduced a parameter $r$
defined by
\begin{eqnarray}
r&\equiv&{F_1^{NF}(m_{\psi'}^2) \over F_1^{NF}(m_{\psi}^2)}
{F_1^{BK}(m_{\psi}^2) \over
F_1^{BK}(m_{\psi'}^2)}, \nonumber \\
&=&{A_1^{NF}(m_{\psi'}^2) \over A_1^{NF}(m_{\psi}^2)} {A_1^{BK^*}(m_{\psi}^2)
\over
A_1^{BK^*}(m_{\psi'}^2)}.
\end{eqnarray}
There is no compelling reason for the equality in eqn. (17); one could have
chosen
independent ratios for $ B \rightarrow \psi (\psi') + K$ and $ B \rightarrow
\psi (\psi') + K^*$
decays and described  the data equally well.

        Clearly, all data for color-suppressed decays, $ B \rightarrow \psi
(\psi') + K(K^*)$, can
be accounted for in all models by using the ``standard'' $ N_c=3$ value for $
a_2(=0.09)$
but with the inclusion of a modest nonfactorized contribution to the amplitude
with the
appropriate Lorentz structure accompanying $F_1^{BK} $ in $ B \rightarrow \psi
K$ and
$A_1^{BK^*} $ in $ B \rightarrow \psi K^*$. In this regard we differ from the
proposal by
Carlson and Milana   \cite{cm} who assume that nonfactorized contributions
effect $
\Gamma_T$ only and not $ \Gamma_L$. In our language it would mean $A_1^{NF}  =
A_2^{NF}  = 0$, $V^{NF} \neq 0$. Our suggestion, $A_1^{NF} \neq 0$, $A_2^{NF}
=
V^{NF}  = 0$, effects both $ \Gamma_T$  and $ \Gamma_L$.

\begin{table}
\begin{center}
\caption{Model predictions, with nonfactorized contribution, for $ P_L$, $R$, $
R'$,
$R^{\prime\prime} $ and $ B(B\rightarrow \psi K)$; $C_1=1.11$ and $C_2=-0.28$,
or,
equivalently $a_1=1.02$ and $a_2=0.09$.
Read $ R'$ and $ R^{\prime\prime}$ with a 12.3\% error and BR with a 7.3\%
error.}
\vspace{3mm}
\begin{tabular}{|c|c|c|c|c|c|c|c|c|}
\hline
&&&&&&&&\\
Model&$ {A_1^{NF} \over A_1^{BK^*}}$&${F_1^{NF} \over F_1^{BK}}$&$r$&$ P_L$&R&$
R^\prime$&$ R^{\prime\prime}$&$BR^{(a)}$\\
&&&&&&&&in $ \%$ \\
\hline
BSWI&0.07&0.23&1.30&0.74&1.48&1.79&1.55&0.094\\
&0.07&0.23&1.40&0.74&1.48&1.59&1.42&0.094\\
\hline
BSWII&0.10&0.13&1.20&0.71&1.98&1.35&1.65&0.095\\
&0.10&0.13&1.30&0.71&1.98&1.21&1.46&0.095\\
\hline
CDDFGN&0.15&0.16&1.30&0.75&1.44&1.86&1.54&0.099\\
&0.15&0.16&1.40&0.75&1.44&1.67&1.38&0.099\\
\hline
HSQ&0.09&0.30&1.30&0.72&1.97&1.26&1.50&0.093\\
&0.09&0.30&1.40&0.72&1.97&1.11&1.36&0.093\\
\hline
JW&0.07&0.23&1.30&0.74&1.48&1.79&1.55&0.094\\
&0.07&0.23&1.40&0.74&1.48&1.59&1.42&0.094\\
\hline
IW&0.11&0.23&1.40&0.76&1.42&1.72&1.57&0.094\\
&0.11&0.23&1.50&0.76&1.42&1.54&1.43&0.094\\
\hline
Expt.&&&&0.78&1.71&1.48&1.13&0.094\\
&&&&$ \pm0.07$&$ \pm0.34$&$ \pm0.46$&$ \pm0.50$&$ \pm0.012$\\
&&&&&&(b)&(b)&\\
\hline
\end{tabular}
\end{center}
(a) BR $\equiv$ B$(B\rightarrow\psi K)$, (b)  Our estimate of error.
\end{table}

           Finally, we show why factorization assumption works so well for all
models in
explaining the polarization data in color-favored  decays. The amplitude
(analogous to  eqn.
(5)) for $ \bar{B}^0 \rightarrow D^{*+} + \rho^-$ decay can be written as

\begin{eqnarray}
A(\bar{B}^0 \rightarrow D^{*+} \rho^-) ={G_F \over \sqrt{2} } V_{cb} V_{ud}^*
a_1\left[
\left\langle D^{*+} \rho^- \mid (\bar{c}b)(\bar{d}u)\mid \bar{B}^0
\right\rangle \right.
\nonumber\\
\left.+  \zeta \left\langle D^{*+} \rho^-\mid \tilde{H}_w^{(8)}\mid \bar{B}^0
\right\rangle\right],
\end{eqnarray}
 where $a_1 = C_1 + C_2 /3 = 1.03\pm0.014$ and $\zeta= C_2/ a_1$ and $
\tilde{H}_w^{(8)}  = {1 \over 2} \sum{(\bar{c}\lambda^au)}$
${(\bar{d}\lambda^ab)}$.  Now,
since $|\zeta| \approx \kappa/40$, it is clear that the role of the
nonfactorized terms is
strongly suppressed compared to the case of color-suppressed decays. As a
consequence,
factorization assumption works well for color-favored decays. Thus assuming
factorization,
$ P_L(\bar{B}^0 \rightarrow D^{*+} \rho^-)$ is given by

\begin{eqnarray}
 P_L(\bar{B}^0 \rightarrow D^{*+} \rho^-) = {(\hat{a} - \hat{b}\hat{x})^2 \over
(\hat{a} -
\hat{b}\hat{x})^2 + 2(1 +\hat{c}^2\hat{y}^2)},
\end{eqnarray}
where the hatted quantities relevant to $\bar{B}^0 \rightarrow D^{*+} \rho^-$
decay are the
analogues of the unhatted ones defined in eqn.  (12). Numerically $ \hat{a}$ is
twice as
large as $ \hat{b}$ and  much larger than $ \hat{c}$: $ \hat{a} = 7.507$, $
\hat{b} = 3.225$
and $ \hat{c} = 0.433$. Thus for $ \hat{x} \approx 1$ and $ \hat{y} \approx
(1-2)$, which
most models predict, the longitudinal polarization is  close to unity, in
agreement with data
\cite{cleo}.

         We wish to emphasize an important difference between $F_1^{BK} $,
$A_1^{BK^*} $
and $F_1^{NF}$, $A_1^{NF}$: Whereas the former, being  form factors,  are
three-point
functions and real at $ q^2 = m_\psi^2$, the latter representing the scattering
of the weak
spurion, $H_w^{(8)}$, $ H_w^{(8)} + B \rightarrow \psi  + K(K^*)$  are
four-point functions. $
F_1^{NF}$ and  $ A_1^{NF}$ are needed at the Mandelstam variables $ s = m_B^2$,
$t =
m_\psi^2 $, $u = m_{K(K^*)}^2$. {\em In general, $ F_1^{NF}$ will be complex
since $ m_B
> (m_\psi + m_K)$.}

In summary, we have proposed that nonfactorized amplitudes play a crucial role
in
color-suppressed $ B\rightarrow\psi(\psi')+K(K^*) $ decays. With the additional
assumption
that the nonfactorized amplitude contribute only to S-wave production in $ B
\rightarrow
\psi(\psi') K^*$ decay, we have demonstrated that all data on $
B\rightarrow\psi(\psi')+K(K^*) $ can be accommodated in the commonly used form
factor
models with the inclusion of a modest nonfactorized contribution.  We emphasize
that
without the nonfactorized contribution,  polarization data in  $ B \rightarrow
\psi  K^*$ decay
cannot be understood  \cite{gkp}.

 For $ B \rightarrow \psi K $ decay one can indeed define an effective $ a_2$,
which could
be complex as $ F_1^{NF}$ would, in general, be complex. This effective $ a_2$
is also
process-dependent.  Despite the ``standard''   $ N_c = 3$ value of $ a_2$ being
$0.10\pm0.03$, since $ \kappa =C_1/a_2$ is of the order of ten, the effective
$a_2$ could
be $ \approx$ 0.22 even for a modest nonfactorized contribution of 10$\%$ in
the amplitude.
For more complex processes involving more than one form factor such as $ B
\rightarrow
\psi K^* $ it is not possible to factor out an effective $ a_2$.

A corollary to our analysis is that the effective

$a_2$ being process-dependent, there is no reason for it to be the same in
color-suppressed $ B \rightarrow \pi(\rho) + D(D^*)$ decays as in $ B
\rightarrow \psi +
K(K^*)$ decays.

Nonfactorized contributions in charmed meson decays were first discussed  by
Deshpande,
Gronau and Sutherland  \cite{ngd}. More recently, Cheng   \cite{hyc}, Cheng and
Tseng
\cite{chng} and Soares  \cite{srs} have used language similar to ours but their
emphasis was
quite different. In ref.  \cite{chng} the authors assume factorization to be
valid and try to
explain the ratios R and $ P_L$ by using modified form factors. However their
predicted $
P_L$ does not satisfy eqn.  (14). The role of nonfactorized contributions to D
and B decays
has  been discussed by  Blok and Shifman (ref. \cite{bshf1,bshf2} and
references therein).
Their emphasis was to understand  the discarding of of the $ 1/N_c$ term in the
definitions $
a_{1,2}=C_{1,2}+C_{2,1}/N_c$. However, if  $ 1/N_c$ terms are discarded $ a_2$
would be
negative, whereas recent experiments  \cite{cleo} leave no doubt that $ a_2$ is
positive.
The prejudice that  $ 1/N_c$ term ought to be discarded (or cancelled by by
nonfactorized
contributions) was carried over to B decays from the experience in D decays,
and the
earlier ARGUS and CLEO data  \cite{nrsx} appeared to support it but such is not
the case at
present.

Finally, our choice $ a_2 = 0.10 \pm0.03$ is consistent with the values $
a_2^{HV} =
0.16\pm0.05 $ and  $ a_2^{DRED} = 0.15\pm0.05 $ quoted by Buras  \cite{burs}
(HV for 't
Hooft-Veltman and DRED for  ``dimensional reduction'', see ref.  \cite{burs}
for details and
references) but is inconsistent with $ a_2^{NDR}=0.20\pm0.05$ (NDR for ``naive
dimensional reduction'').
\vskip 0.5cm

ANK wishes to acknowledge a research grant from the Natural Sciences and
Engineering
Research Council of Canada which partially supported this research.

\newpage
\noindent {\bf Figure Captions}
\vskip 0.5cm
\noindent Fig.1:
The domain of $ (x,y)$ allowed by the polarization data (eqn.  (14)) for $
B\rightarrow \psi
K^*$ with different value of $ \chi (\equiv A_1^{NF}(m_\psi^2) /
A_1^{BK^*}(m_\psi^2))$;
$C_1=1.11$ and $C_2=-0.28$, or, equivalently $a_1=1.02$ and $a_2=0.09$.
Also shown are predictions of various models, A: BSWI, B: BSWII, C: CDDFGN, D:
HSQ, E:
JW and F: IW.


\newpage
\begin{thebibliography}{99}

\bibitem{gkp}
M. Gourdin, A. N. Kamal and X. Y. Pham, University of Paris Report No.
PAR/LPTHE/94-19
(1994) (to be published in Phys. Rev. Lett.)
\bibitem{ks}
A. N. Kamal and A. B. Santra, University of Alberta Report No. Alberta Thy
27-94 (1994) (to
appear in Phys. Rev. D).
\bibitem{cleo}
M. S. Alam et al. ( CLEO collaboration) Phys. Rev. D50  (1994)  43.
\bibitem{pdg}
Particle Data Group, Phys. Rev.  D50  (1994) 1173.
\bibitem{nrsx}
M. Neubert, V. Rieckert, B.Stech and Q. P. Xu, in: Heavy Flavours (World
Scientific,
Singapore, 1992) eds. A. J. Buras and M. Lindner.
\bibitem{bshf1}
B. Blok and M. Shifman, Nucl. Phys. B399  (1993) 441.
\bibitem{sfm}
M. Shifman, Nucl. Phys. B388  (1992)  346;

I. Bigi, B. Blok, M. Shifman,  N. Ultrasev and A. Vainshtein, Report
CERN-TH-7132/94
(1994).
\bibitem{bsw}
M. Bauer, B. Stech and M. Wirbel, Z. Phys. C34  (1987) 103;  M. Wirbel, B.
Stech and M.
Bauer, Z. Phys. C29  (1985)  637.
\bibitem{cdf}
K. Ohl (CDF Collaborations), Invited talk at Division of Particles and Fields
(APS) Meeting,
Albuquerque, N.M., 2-6 August, 1994.
\bibitem{argus}
H. Albrecht (ARGUS Collaboration), DESY Report No. 94-139 (1994).
\bibitem{cddfgn}
R. Casalbuoni, A.Deandrea, N. Di Bartolomeo, R. Gatto, F. Feruglio and G.
Nardulli, Phys.
Lett. B292  (1992)  371;
A. Deandrea, N. Di Bartolomeo, R. Gatto and G. Nardulli, Phys. Lett . B318
(1993)  549.
\bibitem{hsq}
A.Ali, T. Mannel, Phys. Lett. B264 (1991)  447;  B274  (1992)  256;  A. Ali, T.
Ohland and T.
Mannel, Phys. Lett. B298  (1993) 195.
\bibitem{jw}
W. Jaus, Phys. Rev. D41 (1990) 3394; W.Jaus and D. Wyler, Phys. Rev.  D41
(1990) 3405.
\bibitem{iw}
N. Isgur and M. B. Wise, Phys. Rev. D42  (1990) 2388.
\bibitem{cm}
C. E. Carlson and J. Milana, College of William and Mary Report No. WM-94-110
(1994).
\bibitem{ngd}
N. Deshpande, M. Gronau and D. Sutherland, Phys.  Lett. 90B   (1980)  431.
\bibitem{hyc}
H. -Y Cheng, Taipei Report No. IP-ASTP-11-94 (1994).
\bibitem{chng}
H. -Y Cheng and B. Tseng, Taipei Report No. IP-ASTP-21-94 (1994).
\bibitem{srs}
J. M. Soares, TRIUMF Report No. TRI-PP-94-78 (1994).
\bibitem{bshf2}
B. Blok and M. Shifman, Nucl. Phys. B399 (1993) 459;
Nucl. Phys. B389 (1993) 534.
\bibitem{burs}
A. J. Buras, Max-Plank-Institute f\"{u}r Physik Report No. MPI-PhT/94-60
(1994).
\end{thebibliography}
\end{document}